\documentstyle[12pt,aaspp4]{article}
\begin{document}
\title{RR Lyrae stars in Galactic globular clusters.\\
III. Pulsational predictions for metal content $Z$=0.0001 to
$Z$=0.006}
\author{Di Criscienzo, M.\altaffilmark{1,2}, Marconi, M.\altaffilmark{1},
Caputo, F.\altaffilmark{3}} \altaffiltext{1}{INAF - Osservatorio
Astronomico di Capodimonte, via Moiarello 16, I-80131 Napoli,
Italy; marcella@na.astro.it, dicrisci@na.astro.it}
\altaffiltext{3}{INAF - Osservatorio Astronomico di Roma, Via di
Frascati, 33, I-00040 Monte Porzio Catone Italy;
caputo@mporzio.astro.it} \altaffiltext{2}{Universita' "Tor
Vergata", via della Ricerca Scientifica 1, I-00133, Roma, Italy}

\begin{abstract}

The results of nonlinear, convective models of RR Lyrae pulsators
with metal content $Z$=0.0001 to 0.006 are  discussed and several
predicted relations connecting pulsational (period and amplitude
of pulsation) and evolutionary parameters (mass, absolute
magnitude and color of the pulsator) are derived. These relations,
when linked with the average mass of RR Lyrae stars, as
suggested by horizontal branch evolutionary models, provide a
``pulsational'' route to the determination of the distance
modulus, both apparent and intrinsic, of RR Lyrae rich globular
clusters. Based on a preliminary set of synthetic horizontal
branch simulations, we compare the predicted relations with
observed variables in selected globular clusters (M2, M3, M5, M15,
M55, M68, NGC 1851, NGC 3201, NGC 5466, NGC 6362, NGC 6934 and IC
4499). We show that the distance moduli inferred by the various
theoretical relations are mutually consistent within the errors,
provided that the value of the mixing-length parameter slightly
increases from the blue to the red edge of the pulsation region.
 Moreover, we show that the relative ``pulsational'' distance
moduli fit well previous empirical results and that the
 parallax of the prototype variable RR Lyr, as
inferred by the predicted Period-Wasenheit relation, is in close
agreement with the HST astrometric measurement.

\end{abstract}

\keywords{Stars: evolution - stars: horizontal branch - stars:
variables: RR Lyrae}

\section{Introduction}

 RR Lyrae stars represent the most common type of Population
II radially pulsating stars, with approximately 3000 variables in
Galactic globular clusters (see Clement et al. 2001) and
increasing samples being discovered in all Local Group galaxies.
If members of globular clusters (GCs), they show low ($Z\sim$
0.0001) to intermediate metal content ($Z\le$ 0.006), but up to
solar abundances are measured for field variables. They represent
fundamental tracers of ancient stellar populations and, via the
calibration of their absolute visual magnitude $M_V$(RR) in terms
of the measured iron-to-hydrogen content [Fe/H], they are commonly
used as standard candles for distance determinations in the Local
Group, with relevant implications for the cosmic distance scale
and the age of globular clusters (see e.g.  Carretta et al. 2000,
Cacciari 2003). For these reasons, several observational and
theoretical efforts have been made in the last years to provide
accurate evaluations of the $M_V$(RR) versus [Fe/H] relation,
which is commonly approximated in the linear form
$M_V$(RR)=$\alpha$+$\beta$[Fe/H].

Although the discussion of the $M_V$(RR)-[Fe/H] relation is out of
the purpose of the present paper, it is worth mentioning that,
from the side of empirical studies, recent evaluations of the
zero-point $\alpha$ are spanning an interval of about 0.3 mag,
while the value of the slope $\beta$ ranges from $\sim$ 0.13 to
$\sim$ 0.30 mag/dex. As an example, according to the very recent
review by Cacciari (2003), the value of $M_V$(RR) at [Fe/H]=$-$1.5
varies from  0.56$\pm$0.15 mag (``revised'' Baade-Wesselink
method) and 0.62$\pm$0.10 mag (direct astrometry to GCs) to
0.68$\pm$0.08 mag (``classical'' Baade-Wesselink method) and
0.77$\pm$0.13 mag (statistical parallaxes). Concerning the
theoretical approach, we have already discussed (Marconi et al.
2003, hereafter Paper II) that the predicted $M_V$(RR)-[Fe/H]
relation inferred by Horizontal Branch (HB) evolutionary models
depends on several factors (e.g., the color distribution of HB
stars, the adopted ratio between $\alpha$- and heavy elements, the
adopted bolometric correction) and it should be handled with care,
also in consideration of the fact that HB models yield ``static''
magnitudes (i.e. the values the star would have were it not
pulsating), whereas observed magnitudes are ``mean'' quantities
averaged over the pulsation cycle, which not necessarily are
exactly alike static values.

In summary, in spite of a huge amount of work, we are presently
facing a quite intriguing scenario: on the evolutionary side, even
for a fixed dependence of the luminosity $L(RR)$ on $Z$, the
predicted $M_V$(RR)-[Fe/H] relation turns out to depend on several
assumptions (bolometric correction, [$\alpha$/Fe] ratio,
evolutionary status of the variables), while on the observational
side  no firm evaluation has been reached yet. Note that even the
recent HST determination of a quite accurate trigonometric
parallax ($\pi_{trig}$=3.82$\pm$0.20 mas) for the prototype
variable RR Lyr ([Fe/H]=$-$1.39) yields  $M_V$=0.607$\pm$0.114
mag$\pm$0.034 mag, where the last two figures illustrate,
respectively, the intrinsic uncertainty of the HST measurement and
the effect of an interstellar extinction $A(V)$=0.087$\pm$0.034
mag (see Benedict et al. 2002, Bono et al. 2003).

In this context, the study of quantities such as the pulsation
period and amplitude is of fundamental importance since they
depend on the star structural parameters (mass, luminosity and
effective temperature) and are distance and reddening free
observable quantities, as well. In order to perform a thorough
analysis of observed RR Lyrae stars (see Castellani, Caputo \&
Castellani 2003, Paper I, for the observational scenario of these
variables in Galactic globular clusters), we planned to construct
a theoretical framework based on both pulsation and evolution
models computed at various metal content and using the same input
physics. The general procedure has been discussed in detail in
Paper II, where the results at $Z$=0.001, the typical metal
content of Oosterhoff type I clusters, are compared with RR Lyrae
stars in M3. In the present paper,  we discuss nonlinear
convective pulsation models for $Z$=0.0001, 0.0004 and 0.006,
while an extensive atlas of synthetic HB computations will be
given in a future paper (Cassisi, Castellani, Caputo \& Castellani
2004, Paper IV, in preparation).

Following the organization of Paper II, we present in Section 2
the pulsation models and the relevant relations among the various
parameters, while a first comparison with observed variables in RR
Lyrae rich globular clusters is given in Section 3. Some brief
remarks close the paper.

\section{The theoretical pulsational scenario}

\subsection{Periods, magnitudes and colors}

 The existing grid of nonlinear and convective models, as
computed with a mixing length parameter $l/H_p$=1.5 (see Bono et
al. 2001, 2003 and Paper II), has been extended to new fundamental
(F) and first-overtone (FO) computations with $Z$=0.0001, adopting
mass $M$=0.80$M_{\odot}$ and log$L/L_{\odot}$=1.72, 1.81, 1.91 (Di
Criscienzo 2002) and $M$=0.65$M_{\odot}$ and log$L/L_{\odot}$=1.61
(present paper) . As a result, we are dealing with the set of
pulsation models listed in Table 1. The bolometric light curves of
all these models have been transformed into the observational
plane by adopting the bolometric corrections and temperature-color
transformations provided by Castelli, Gratton \& Kurucz (1997a,b,
hereafter CGKa,b). In this way, light curve amplitudes and mean
absolute magnitudes, either intensity-weighted $\langle
M_i\rangle$ or magnitude-weighted $(M_i)$, are derived in the
various photometric bands.

As a first step we investigated the dependence, if any, of the
pulsation period relation $P=f(M,L,T_e)$ on the metal content.
After confirming that the period of FO pulsators can be
fundamentalised adopting log$P_F$=log$P_{FO}$+0.130$\pm$0.004, we
derive (see Fig. 1) that all over the explored range of mass,
luminosity and metal content, the pulsation equation can be
approximated as

$$\log P_F=11.038(\pm0.003)+0.833(\pm0.003)\log L/L_{\odot}-0.651(\pm0.007)\log M/M_{\odot}$$
$$-3.350(\pm0.009)\log T_e+0.008(\pm0.001)\log Z,\eqno(1)$$

\noindent in close agreement with earlier results presented by
Caputo, Santolamazza \& Marconi (1998).

As for the limits of the  whole instability strip, namely the
effective temperature of the first-overtone blue edge (FOBE) and
of the fundamental red edge (FRE), we find no significative
metallicity effect on the FOBE relation at $Z$=0.001, as given in
Paper II, whereas the FRE tends to be slightly redder, for fixed
mass and luminosity, when increasing the metal content. However,
as already discussed in Paper II, one should also consider that
varying the value of the mixing length parameter $l/H_p$ will
modify the edges of the  pulsation region, shrinking the width of
the whole instability strip if increasing the efficiency of
convection in the star external layers. In order to study this
effect, all the models listed in Table 1, but the case $Z$=0.0004,
have been recomputed by adopting $l/H_p$=2.0. The results confirm
that the FOBE effective temperature decreases by $\sim$ 100 K,
whereas the FRE one increases by $\sim$ 300 K. Moreover, we notice
that for the same $l/H_p$ variation the fundamental blue edge
changes by 100 K at most, while the first overtone red edge gets
hotter by about 200 K.

On this ground, the FOBE and FRE relations at $l/H_p$=1.5 given in
Paper II can be retained over the whole range $Z$=0.0001-0.006,
with a total uncertainty of $\pm$ 0.005 and 0.010, respectively,
while the effects of a variation of $l/H_p$ from 1.5 to 2.0 on the
edges of the instability strip can be quantified as
$\Delta$log$T_e^{FOBE}$= $-$0.008$\pm$0.003 and
$\Delta$log$T_e^{FRE}$=0.021$\pm$0.005. However, it is of interest
to note that with $Z$=0.006 and log$L/L_{\odot}\ge$ 1.65 the FO
instability strip disappears when $l/H_p$=2.0 is adopted. Such a
result, together with the observational evidence of $c$-type RR
Lyrae stars with such a metal content, seems to suggest that
$l/H_p$=2.0 is a too high value for metal-rich pulsational models.

 Concerning the location in the $M_V$-log$P$ diagram of the
instability strip theoretical edges, taking also into account the
above mentioned effects of the mixing-length parameter, a
linear regression to the models yields

$$M_V^{FOBE}=-1.19-2.23\log M/M_{\odot}-2.54\log P_{FO}+0.10(l/H_p-1.5)\eqno(2)$$

\noindent and

$$M_V^{FRE}=0.12-1.88\log M/M_{\odot}-1.96\log P_F-0.28(l/H_p-1.5)\eqno(3)$$

\noindent with a standard deviation of 0.05 and 0.06 mag,
respectively.

Before proceeding, let us first recall that, when dealing with
pulsating structures, the static magnitudes differ from the
measured magnitudes, which are mean quantities (intensity-weighted
or magnitude-weighted) averaged over the pulsation cycle. This
issue has been discussed in detail in Paper II, and here we wish
only to mention that the results at $Z$=0.001 hold all over the
range $Z$=0.0001-0.006. In particular, we find that the
discrepancy between static and mean values is a function of the
pulsation amplitude, decreasing from asymmetric to symmetric light
curves as well as when passing from optical ($BVI$) to
near-infrared ($K$) bands, and that intensity averaged quantities
are better representative of static magnitudes and colors. As a
whole, each mean magnitude can be corrected for the amplitude
effect to get the corresponding static value, or, alternatively,
static magnitudes can be directly estimated when the two types of
mean values are measured. However, since in the present paper we
aim at providing a handy tool for the analysis of observed data,
all the following relations will be given for intensity-averaged
magnitudes\footnote{The whole set of computed models are available
upon request to marcella@na.astro.it.}.

 The obvious outcome of the period relation [see equation (1)]
into the observational plane is the Period-Magnitude-Color ($PMC$)
relation, where the pulsation period for each given mass is
correlated with the pulsator absolute magnitude and color. 
When using linear pulsating models, this could be made by simply
transforming luminosity and effective temperature into absolute
magnitude and color, but the result is a {\it static} $PMC$
relation which cannot be directly compared  to observed variables.
On the contrary, our non linear approach supplies {\it mean}
magnitudes and colors. Based on intensity-averaged magnitudes, the
linear interpolation through the results (adopting fundamentalised
periods for FO models) gives for each adopted metal content $Z$

$$\log P'_F=\log P_F+0.34\langle M_V\rangle+0.54\log M/M_{\odot}=$$
$$a(Z)+b(Z)[\langle M_B\rangle -\langle M_V\rangle]$$

\noindent where the coefficients $a(Z)$ and $b(Z)$ are listed in
Table 2.

As discussed in Paper II, for fixed mass and metal content, the
$PMC$ relation defines accurately the properties of individual
pulsators within the instability strip, but its use to get
distances needs accurate reddening evaluations. However, it is of
interest to mention that the magnitude dispersion due to the
finite width of the instability strip is rather close to the
effect of interstellar extinction. Based on such an evidence, the
Wesenheit functions (see Dickens \& Saunders 1965)
$WBV$=$V-3.1(B-V)$, $WVI$=$V-2.54(V-I)$, etc., which have been
widely used in Cepheid studies (see  Madore 1982 and  Madore \&
Freedman 1991), can be adopted also for RR Lyrae stars (see Kovacs
\& Jurcsik 1997, Kovacs \& Walker 2001) to  provide a
reddening-free $PW$ relation, where also the dispersion of
magnitude at a given period is significantly reduced. Before
proceeding, let us remark the deep difference between the $PMC$
and $PW$ relations: the former one  is defined by the
intrinsic properties of the variables and cancels the effects of
the finite width of the instability strip, while the latter 
depends on the properties of the interstellar medium and removes
any reddening effect, differential or total. Therefore,  in the
case of variables at the same distance and with the same mass and
metal content, the scatter in observed $PMC$ relations, excluding
photometric errors, should  mainly depend on reddening,
whereas the scatter in observed $PW$ relations is a residual
effect of the finite width of the strip.

 Using $\langle WBV\rangle$ quantities based on
intensity-averaged magnitudes, a linear interpolation through F
and fundamentalised FO models yields the Period-Wesenheit $PWBV$
relation

$$\langle WBV\rangle=-1.655(\pm0.054)-2.737(\pm0.027)\log P^W_F\eqno(4a)$$

\noindent where

$$\log P^W_F=\log P_F+0.54\log M/M_{\odot}+0.03\log Z$$

\noindent As shown in Fig. 2, the correlation is quite tight,
providing a useful way to get the intrinsic distance modulus of
variables with known mass and metal content with a formal accuracy
of 0.06 mag (dashed lines).

As for the $\langle WVI\rangle$ function, the linear interpolation
through F and fundamentalised FO models yields

$$\langle WVI\rangle=-1.670(\pm0.030)-2.750(\pm0.013)\log P^W_F\eqno(4b)$$

\noindent where

$$\log P^W_F=\log P_F+0.65\log M/M_{\odot}$$

 It is of interest to note that RR Lyrae stars observed in the
Large Magellanic Cloud yield $\delta \langle WVI\rangle/\delta$
log$P_F\sim -$2.75 (Soszynski et al. 2003) which is the predicted
slope at constant mass. Moreover, variables observed in Galactic
globular clusters suggest $\delta (WBV)/\delta$log$P_F\sim -$2.47
and $\delta (WVI)/\delta$log$P_F\sim -$2.51 (see Kovacs \& Walker
2001), where $(WBV)$ and $(WVI)$ are magnitude-averaged
quantities. Also these results appear in reasonable agreement with
the predicted slopes we get at roughly constant mass and metal
content, namely $\delta (WBV)/\delta$log$P_F=-2.66(\pm0.03)$ and
$\delta (WVI)/\delta$log$P_F=-2.64(\pm0.02)$.

In matter of distance determinations, let us finally consider the
predicted near-infrared Period-Magnitude ($PM_K$) relation, which
is well known to be a powerful tool to get distances given the
quite low dependence of the $K$-magnitude on interstellar
extinction. Adopting again intensity-averaged magnitudes (but in
the $K$-band the difference between intensity-averaged,
magnitude-averaged and static values is quite small due to the
reduced pulsation amplitudes), we show in Fig. 3 that  linear
regression through F and fundamentalised FO pulsators provides:

$$\langle M_K\rangle=-0.094-2.158\log P_F-1.436\log M/M_{\odot}-0.712\log L/L_{\odot}\eqno(5)$$
\noindent  with a standard deviation of $\pm$ 0.04 mag (dashed
lines).

\subsection{Pulsation amplitudes}
Figures 4 and 5  show bolometric light curves with $l/H_p$=1.5
(solid line) and $l/H_p$=2.0 (dashed line) for a number of F and
FO models with $M=0.80{M_{\odot}}$ at varying luminosity and
effective temperature. In agreement with the results presented in
previous papers (Bono \& Stellingwerf 1994, Bono et al. 1997a),
the morphological features of the predicted light curves appear a
function of luminosity and effective temperature, thus providing a
useful tool to constrain such fundamental intrinsic parameters.
 In particular, Bono \& Stellingwerf (1994) have already
discussed that moving toward lower luminosity levels, for a fixed
mass, the morphology of the light curve of FO models becomes very
similar to that of fundamental pulsators, changing from almost
sinusoidal to sawtooth and with a significant increase of the
bolometric amplitude. In this context, the use of the Fourier
parameters of observed light curves (see, e.g., Kovacs \& Walker
2001 and references therein) appears as a promising way to
approach the determination of the intrinsic parameters, even
though the obvious ultimate goal to take into account all the
observed features (e.g., bump, double-peak) is the best fit of the
whole light curve, as recently presented by Bono, Castellani \&
Marconi (2000, 2002), Castellani, Degl'Innocenti \& Marconi
(2002), Di Criscienzo, Marconi \& Caputo (2003).

Here, as a broad outline of the pulsational behavior, let us
consider the quite linear correlation between fundamental
bolometric amplitudes and periods (logarithmic scale), for fixed
mass and luminosity. However, let us remind that the pulsation
amplitudes, while sharing with the periods the property of being
sound measurements independent of reddening and distance
uncertainties, do significantly depend on the adopted efficiency
of the convective transfer.  As an example, we list in Table 3
the computed periods, colors and visual amplitudes for fundamental
pulsators with $Z$=0.001, 0.75$M_{\odot}$, log$L/L_{\odot}$=1.61
and $l/H_p$=1.5 and 2.0. One has that both the period and the
color are almost independent of the adopted value for the
mixing-length parameter, whereas the amplitude decreases with
increasing the efficiency of convection, for fixed effective
temperature. Moreover, given the already mentioned effects of the
same $l/H_p$ variation on the fundamental red edge (where the
amplitude reaches its minimum value) and blue edge (where the
amplitude reaches its maximum value), the $A_V-$log$T_e$ slope
becomes steeper with increasing the $l/H_p$ value.

As a whole,  after transforming all the bolometric light curves
into the observational plane,  we show in Fig. 6 that all over the
explored metallicity range the Period-Magnitude-Amplitude
($PMA_V$) fundamental relation can be approximated as

$$\log P^*_F=\log P_F+0.385\langle M_V\rangle+0.30\log M/M_{\odot}=0.13-0.189A_V\eqno(6)$$

\noindent at $l/H_p$=1.5, and

$$\log P^*_F=\log P_F+0.385\langle M_V\rangle+0.35\log M/M_{\odot}=0.03-0.142A_V\eqno(7)$$

\noindent at $l/H_p$=2.0, with a standard deviation (dashed lines)
of 0.04 and 0.03 mag, respectively.

Closing this section devoted to pulsation amplitudes,  we
notice that increasing the $l/H_p$ value yields smaller amplitudes
for fixed $BV$ color (see data in Table 3). This is shown in Fig.
7, where the visual amplitude $A_V$ is plotted as a function of
the intensity-weighted $\langle M_B\rangle-\langle M_V\rangle$
color for F pulsators with $l/H_p$=1.5 (upper panel) and
$l/H_p$=2.0 (lower panel). At variance with the results derived
with static colors, here one derives that the 
amplitude-color correlation is not strictly linear as a
consequence of the fact that at larger amplitudes the
intensity-averaged color becomes bluer than the static one (see
Paper II). As a whole, in the explored range of mass and
luminosity, all the results with $l/H_p$=1.5 and $Z\le$ 0.001
suggest the following Color-Amplitude ($CA$) relation

$$\langle M_B\rangle-\langle M_V\rangle=0.469-0.077A_V-0.044A_V^2\eqno(8a)$$
\noindent with a standard deviation of 0.04 mag, while at
$Z$=0.006 we get

$$\langle M_B\rangle-\langle M_V\rangle=0.568-0.156A_V-0.033A_V^2\eqno(8b)$$
\noindent with a standard deviation of 0.02 mag. As for the case
$l/H_p$=2.0, the regression through all the models yields

$$\langle M_B\rangle-\langle M_V\rangle=0.465-0.079A_V-0.035A_V^2+0.014\log Z\eqno(9)$$
\noindent with a standard deviation of 0.03 mag.

\section{Comparison with observed RR Lyrae stars}

As discussed in Paper II, synthetic horizontal branch (SHB)
simulations based on HB evolutionary tracks and pulsational
constraints provide reliable information on the HB morphology (as
given by the ratio (B-R)/(B+V+R), where B, V and R represent the
numbers of blue, variable and red HB stars), as well as on the
``evolutionary'' FOBE and FRE of the synthetic pulsator
distribution, for each given metal content and average mass of HB
stars. It follows that, known the metal content and HB morphology,
one can evaluate the average mass of RR Lyrae stars to be inserted
into the mass-dependent relations presented in the previous
section. As an example, we show in Fig. 8 the correlation between
the HB morphology and the average mass $\langle M(RR)\rangle$
(solar units) of RR Lyrae stars, as derived from synthetic
simulations based on HB models with $Z$=0.0001, 0.0003, 0.0006,
0.001, 0.003 and 0.006 (Paper IV, in preparation). In this figure,
the large dots depict three selected clusters (M15, M3 and NGC
6362) which are plotted according to their observed HB type and
measured [Fe/H] value (see Table 4), under the assumption of
solar-scaled ([$\alpha$/Fe]=0, filled dots) or $\alpha$-enriched
([$\alpha$/Fe]$\sim$0.5, open dots) chemical compositions.

In view of the somehow different HB models published in the recent
literature, the present paper was mainly intended to illustrate
the results of pulsation models, leaving the mass as a free
parameter and  reserving to Paper IV the connection to
stellar evolution theory and the construction of synthetic
relations based also on the evolutionary times. Nevertheless, in
order to check the internal consistency of the pulsational
framework, as well as its appropriateness at the various metal
contents, let us rely on the results plotted in Fig. 8 to compare
the predictions presented in the previous section with observed
data of RR Lyrae rich globular clusters with available $B,V,A_V$
measurements. The selected clusters are listed in Table 4,
together with their [Fe/H] value (Kraft \& Ivans 2003), HB type,
apparent distance modulus $\mu_V$ and $E(B-V)$ reddening (from
Harris 1996), and the source of RR Lyrae data. The last two
columns in Table 4 give the overall metal content $Z$ for a
solar-scaled chemical composition (i.e., log$Z$=[Fe/H]$-$1.70) and
the average mass of the RR Lyrae stars, as inferred from the
results plotted in Fig. 8.

 A general overview of data is shown in Fig. 9, where the variables  are plotted in the
color-magnitude diagram according to intensity-averaged
magnitudes, scaled to an arbitrary value $V_r$, and intrinsic
colors, as derived using $E(B-V)$ values in Table 4. Filled and
open dots are $RR_{ab}$ and $RR_c$ stars, as given by the authors,
and some outliers with respect to the average distribution are
marked.

Starting with the $PW$ relation, which is independent of the
adopted mixing-length parameter, we show in Fig. 10 the comparison
between observed $\langle WBV\rangle$ quantities (symbols as in
Fig. 9) and the predicted relation [see equation (4a)]. For each
cluster, log$P^W_F$ is evaluated according to the mass and metal
content listed in Table 4, while the solid lines drawn in the
panels depict the predicted behavior at constant mass and
metallicity, as reasonably adopted for RR Lyrae stars in a given
globular cluster. In this reddening-free diagram, only the NGC
6934 variables exhibit a considerable scatter, suggesting that
observational errors are possibly occurring in the data. For the
remaining clusters, with the exception of very few outliers, the
observed data fit quite well the predicted slope, allowing us to
derive the intrinsic distance moduli labelled in the various
panels of the figure and listed in column (2) of Table 5. It is
quite interesting to note that these theoretical distance moduli
appear in agreement with previous empirical determinations. This
is shown in Fig. 11, where the relative values with respect to IC
4499, taken as reference cluster, are plotted versus the
evaluations given by Kovacs \& Walker (2001).

Before proceeding with further comparison between theory and
observation, one has to notice that all other predictions depend
on the adopted mixing-length parameter. One may attempt to have a
look on such an issue making use of the predicted Color-Amplitude
fundamental relation, which is independent of the pulsator mass,
and of the reddening values given in Table 4. Figure 12 shows the
visual amplitude of $RR_{ab}$ stars in our sample versus the
intrinsic color $[\langle B\rangle-\langle V\rangle]_0$, in
comparison with the theoretical relations at $l/H_p$=1.5 (equation
(8a): solid line) and 2.0 (equation (9): dashed line).
Disregarding M5 and M55, for the scarcity of data, and NGC 6934,
for the well known significantly inhomogeneous reddening, one
finds that all the remaining clusters seem to exclude $l/H_p$=1.5,
except NGC 3201 and NGC 5466. On this basis, even if tempted to
take $l/H_p$=2.0 as the preferred choice, we keep both the
alternative values and we use equations (8a) and (9) to derive the
$E(B-V)$ values listed in columns (3) and (4) of Table 5.

Adopting again the mass values in Table 4, we use now the
predicted $PMA$ relation [see equations (6) and (7)] for
fundamental pulsators to estimate the apparent distance modulus.
Figure 13 shows the $RR_{ab}$ variables plotted in the
$A_V$-log$P^*_F$ plane, in comparison with the predicted behavior
at constant mass and $l/H_p$=1.5 (solid line). The derived
apparent distance moduli are labelled in the various panels and
listed in column (5) in Table 5, while the following column in the
same table gives the results of a similar comparison, but with
$l/H_p$=2.0.

Finally, we show in Fig. 14 the comparison between the observed
distribution of RR Lyrae stars in the $\langle V\rangle$-log$P$
plane and the predicted edges of the instability strip, as
estimated inserting the mass values in Table 4 into equations (2)
and (3). We start fitting the predicted FOBE at $l/H_p$=1.5 (solid
line at the left) to the observed distribution of $c$-type
variables under the condition that no $c$-type variables are left
in the hot stable region. As shown in Fig. 14, the apparent
distance modulus derived in such a way (the values are labelled in
the panels and listed in column (2) of Table 6) yields a quite
good agreement between the $ab$-type distribution and the
predicted FRE (solid line at the right), except for M68, M15, NGC
5466, M55 and NGC3201.  For these clusters, the observed period
distribution of $ab$-type variables would suggest a red edge as
depicted by the dashed line. In other words, under the assumption
that the whole instability strip is populated, for these clusters
the right value of the distance modulus to fit the observed
distribution of $ab$-type variables is smaller by $\sim$ 0.15 mag
than the value derived from the FOBE-procedure (see column (3) in
Table 6).

Let us now repeat the comparison, but with $l/H_p$=2.0. Since the
predicted FOBE becomes fainter by $\sim$ 0.05 mag, the right
distance modulus to fit the observed distribution of $c$-type
variables decreases by the same quantity. However, at the same
time the predicted FRE becomes brighter by $\sim$ 0.14 mag and, as
shown in Fig. 15, for some clusters several $ab$-type variables
would have longer periods than the predicted FRE. Specifically,
for IC 4499, NGC 6934, M2, M3, M5, NGC1851 and NGC6362 the
comparison of the predicted FRE to the observed red edge (dashed
line) would yield now a distance modulus which is larger by $\sim$
0.10 mag than the one derived from the FOBE-procedure.

We summarize in Table 6 the distance moduli inferred from the
$M_V$-log$P$ distribution. One can conclude that the assumption
$l/H_p$=1.5 yields rather discordant results for M68, M15, NGC
5466, M55, and NGC3201, where a quite close agreement is present
with $l/H_p$=2.0. On the other hand, the assumption $l/H_p$=2.0
yields rather discordant results for the remaining clusters (IC
4499, NGC 6934, M2, M3, M5, NGC1851 and NGC6362), for which a
better agreement is reached with $l/H_p$=1.5. In summary, present
results would suggest that the assumption of a constant
mixing-length parameter should be revisited.

However, recent evolutionary studies seem to exclude any trend of
the $l/H_p$ ratio with the metal content (see Palmieri et al.
2002, Lastennet et al. 2003). Conversely, specific tests aiming at
fitting the light curves of observed pulsators (both RR Lyrae
stars and Classic Cepheids) with modeled curves have shown that
first overtone  (hotter pulsators) light curves are well
reproduced with $l/H_p$=1.5, whereas in the case of fundamental
variables (cooler pulsators) a value $l/H_p\ge$ 1.8 is needed in
order to properly reproduce both the amplitude and the morphology
of observed curves (see Bono, Castellani \& Marconi 2000, 2002;
Castellani, Degl'Innocenti \& Marconi 2002; Di Criscienzo,
Marconi, Caputo 2003).

 According to such  a suggestion of a mixing-length parameter
increasing from the FOBE to the FRE (see also Paper II), we
eventually list in column (6) of Table 6 the apparent distance
modulus of each cluster, as given by the weighted mean over the
values listed in column (6) of Table 5 and in column (2) and
column (5) of Table 6. As shown in Fig. 16, the comparison between
our relative distance moduli and those obtained by Kovacs \&
Walker (2001) on the basis of empirical relations based on Fourier
parameters of the light curves discloses again a quite good
agreement. As for the absolute values, let us remark that the
distance moduli reported in Table 6 hold adopting the SHB
simulations depicted in Fig. 8, CGKa and CGKb bolometric
corrections, as well as solar-scaled chemical compositions.
Concerning this last point, the same Fig. 8 shows the decrease of
the RR Lyrae average mass ($\Delta$log$M\sim-$0.03) if adopting
$\alpha$-enriched ([$\alpha$/Fe]$\sim$0.5) chemical mixtures,
namely $\Delta$log$Z\sim$0.38 for any fixed [Fe/H] value. As a
consequence, the values listed in Table 5 and Table 6 should be
decreased by $\Delta\mu_V$(FOBE/FRE)$\sim -$0.06 mag,
$\Delta\mu_V(PMA)\sim -$0.03 mag and $\Delta\mu_0(PW)\sim -$0.01
mag.

\section{Conclusions}

It is widely recognized that the main tool for interpreting the
observed properties of HB stars is based on the constraints
predicted by synthetic horizontal branches. Such synthetic
simulations require updated evolutionary models and, in order to
extend the analysis to RR Lyrae variables, trustworthy pulsating
models computed with the same updated input physics (e.g. opacity
tables) and assumptions on free parameters (e.g. $l/H_p$).

In Paper II we have shown that such an approach yields a
pulsational route to accurate distance determinations for the
globular cluster M3. The main purpose of the present paper is to
complete the pulsational study,  providing theoretical relations
for the analysis of RR Lyrae stars with metal content $Z$=0.0001
to 0.006.

The comparison with variables in selected globular clusters shows
that the slope of the predicted relations is in reasonable
agreement with the observed distributions, allowing us to proceed
in determining the true distance modulus, as well as reddening and
apparent distance moduli under two assumptions for the
mixing-length parameter. Adopting recent suggestions that the
value of such a parameter may increase from the blue to the red
edge of the pulsation region, we obtain a close agreement among
the distance moduli inferred by the various approaches, supporting
the self-consistency of the pulsational scenario. Moreover, we
show that the relative distance moduli derived in this paper, both
apparent and intrinsic, match well those inferred by empirical
relations (see Kovacs \& Walker 2001).

As for the absolute values, we emphasize that they depend on the
adopted evolutionary mass of the variables which, in turn, mainly
depends on the adopted ratio between $\alpha$ and heavy elements,
decreasing when passing from solar-scaled to $\alpha$-enriched
mixtures, for fixed [Fe/H]. On this issue, it is worth mentioning
that empirical results suggest [$\alpha$/Fe]$\sim$ 0.3 (Carney
1996, Lee \& Carney 2002) and that such a value is consistent with
the global features of color-magnitude diagrams (Caputo \& Cassisi
2002). Finally, let us remind that different bolometric
corrections are still used by the various researchers (see Paper
II) and that the current discrepancy of $\sim$ 0.05-0.06 mag
requires some firm solutions.

However, it is of interest to note that a previous comparison (see
Bono et al. 2003) between pulsational prescriptions and observed
$V$ and $K$ magnitudes of the prototype variable RR Lyr yields a
``pulsational'' parallax quite close with the HST direct
measurement. Using here magnitude-averaged $(B)$ and $(V)$ data,
as listed in Bono et al. (2003), we get that the observed
magnitude-averaged Wasenheit function of RR Lyr is $(WBV)$=6.634
mag. In order to transform this figure into the intensity-weighted
value, we use the predicted correlation plotted in Fig. 17, where
$\Delta WBV$=$\langle WBV\rangle-(WBV)$ and $A_V$ is the visual
amplitude. With $A_V$(RR Lyr)=0.89 mag, one derives $\langle
WBV\rangle$=6.68$\pm$0.02 mag, where the error is due to the
intrinsic spread in Fig. 17. Adopting for RR Lyr, log$P$=$-$0.2466
and [Fe/H]=$-$1.39 (see Bono et al. 2003 for references), and
assuming a mass in the range of 0.66 (as adopted in M5 with
[Fe/H]=$-$1.29) to 0.69$M_{\odot}$ (as adopted in M3, with
[Fe/H]=$-$1.50), equation (4a) gives the absolute value $\langle
W(BV)\rangle$=$-$0.47$\pm$0.05 mag, which means an intrinsic
distance modulus $\mu_0$=7.15$\pm$0.06 mag and a ``pulsational''
parallax $\pi_{puls}$=3.72$\pm$0.10 mas, which is consistent with
the HST geometric measurement $\pi_{trig}$=3.82$\pm$0.20 mas.

\section{Acknowledgments}

It is a pleasure to thank G. Bono, V. Castellani and A. Walker for
a critical reading of the manuscript and useful comments and
suggestions. We thank A. Walker for the data on NGC 6362 and Marco
Castellani for making available synthetic HB simulations. Finally,
we thank our referee for very useful comments. Financial support
for this work was provided by
 the scientific project ``Stellar Populations in Local Group Galaxies''
(P.I. Monica Tosi) from MIUR-Cofin 2002, and by ``Continuity and
Discontinuity in the Milky Way Formation'' (P.I Raffaele Gratton)
from MIUR-Cofin 2003. Model computations made use of resources
granted by the "Consorzio di Ricerca del Gran Sasso", according to
Project 6 "Calcolo Evoluto e sue Applicazioni (RSV6) - Cluster
C11/B".

\pagebreak

\clearpage
\clearpage

\begin{table}[h]
\begin{center}
\caption{Stellar parameters for the full set of pulsation models.
All the computations, except those with $Z$=0.0004, adopt
$l/H_p$=1.5 and 2.0.}
 \vspace{0.3 cm}
\begin{tabular}{llcc}
\hline
 Y & Z & M/M$_{\odot}$ & logL/L$_{\odot}$ \\
\hline
0.24 & 0.0001  &  0.80 & 1.72, 1.81, 1.91\\
       &            &  0.75 & 1.61, 1.72, 1.81\\
       &            &  0.70 & 1.72            \\
        &           &  0.65 & 1.61            \\
\hline
0.24 & 0.0004  &  0.70 & 1.61, 1.72, 1.81\\
\hline
0.24  & 0.001   &  0.75 & 1.71            \\
        &           &  0.65 & 1.51, 1.61, 1.72\\
\hline
0.255 & 0.006   &  0.58 & 1.55, 1.65, 1.75\\
\hline
\end{tabular}
\end{center}
\end{table}

\begin{table}[h]
\begin{center}
\caption{Predicted Period-Magnitude-Color relation
 log$P'_F$=a+b[$\langle M_B\rangle-\langle M_V\rangle$] with
intensity-averaged $BV$ magnitudes.} \vspace{0.3 cm}
\begin{tabular}{lcc}
\hline
Z & a & b \\
\hline
0.0001  &  $-$0.552 $\pm$ 0.014 &  1.290 $\pm$ 0.029\\
0.0004  &  $-$0.543 $\pm$ 0.014 &  1.256 $\pm$ 0.029\\
0.001   &  $-$0.545 $\pm$ 0.014 &  1.208 $\pm$ 0.028\\
0.006   &  $-$0.522 $\pm$ 0.018 &  1.019 $\pm$ 0.029\\
\hline
\end{tabular}
\end{center}
\end{table}

\begin{table}[h]
\begin{center}
\caption{Effects of the mixing-length parameter $l/H_p$ on the
period, color and visual amplitude of fundamental models with
$Z$=0.001, 0.75$M_{\odot}$ and log$L/L_{\odot}$=1.61.}
\begin{tabular}{cccccccc}
\hline $T_e$(K) & $P$(days) & $\langle M_B\rangle-\langle
M_V\rangle$ & $A_V$ && $P$(days) & $\langle M_B\rangle -\langle
M_V\rangle$
& $A_V$\\
\hline
 &  & ($l/H_p$=1.5) && & & ($l/H_p$=2.0) &\\
6100 & 0.5697 & 0.423 & 0.517 && stable & -- & --\\
6200 & 0.5400 & 0.405 & 0.748 && stable & -- & --\\
6300 & 0.5120 & 0.385 & 0.813 && 0.5109 & 0.379 & 0.154 \\
6400 & 0.4859 & 0.362 & 0.922 && 0.4854 & 0.362 & 0.548 \\
6500 & 0.4622 & 0.337 & 1.162 && 0.4604 & 0.342 & 0.649 \\
6600 & 0.4385 & 0.310 & 1.240 && 0.4384 & 0.320 & 0.851 \\
6700 & 0.4170 & 0.285 & 1.413 && 0.4164 & 0.295 & 1.074 \\
6800 & 0.3969 & 0.263 & 1.544 && 0.3966 & 0.271 & 1.260 \\
6900 & 0.3779 & 0.244 & 1.615 && 0.3780 & 0.250 & 1.385 \\
7000 & FO     & --    & --    && 0.3608 & 0.231 & 1.416 \\
\hline
\end{tabular}
\end{center}
\end{table}

\begin{table}[h]
\begin{center}
\caption{Selected RR Lyrae rich Galactic globular clusters listed
with their [Fe/H] value (Kraft \& Ivans 2003), HB type, apparent
distance modulus $\mu_V$ and $E(B-V)$ reddening (Harris 1996). The
reference of RR Lyrae data is also given. The last two columns
give the overall metal content $Z$, adopting a solar-scaled
chemical composition, and the average mass $\langle M(RR)\rangle$
of RR Lyrae stars, as estimated by synthetic horizontal branch
simulations (see text).} \vspace{0.3 cm}
\begin{tabular}{lccccccc}
\hline NGC/IC &[Fe/H] &HB &$\mu_V$ &$E(B-V)$ &Ref. (RR Lyrae)
&log$Z$
&$\langle M(RR)\rangle$\\
   &      &  & (mag) & (mag)  &          & [$\alpha$/Fe=0] & $M/M_{\odot}$ \\
\hline
4590-M68 & $-$2.43 &   0.44 & 15.19 & 0.05 & Wa94      & $-$4.13 & 0.80\\
7078-M15 & $-$2.42 &   0.67 & 15.37 & 0.10 & SS95      & $-$4.12 & 0.77\\
5466     & $-$2.22 &   0.58 & 16.00 & 0.00 & Co99      & $-$3.92 & 0.74\\
6809-M55 & $-$1.85 &   0.87 & 13.87 & 0.08 & Ol99      & $-$3.55 & 0.69\\
4499*    & $-$1.60 &   0.11 & 17.09 & 0.23 & Wa96      & $-$3.30 & 0.70\\
6934     & $-$1.59 &   0.25 & 16.29 & 0.10 & Ka01      & $-$3.20 & 0.70\\
7089-M2  & $-$1.56 &   0.96 & 15.49 & 0.06 & LC99      & $-$3.26 & 0.66\\
3201     & $-$1.56 &   0.08 & 14.21 & 0.23 & Pi02      & $-$3.26 & 0.69\\
5272-M3  & $-$1.50 &   0.08 & 15.12 & 0.01 & CC01      & $-$3.20 & 0.69\\
5904-M5  & $-$1.26 &   0.31 & 14.46 & 0.03 & Ka00+Ca99 & $-$2.96 & 0.66\\
1851     & $-$1.19 &$-$0.36 & 15.47 & 0.02 & Wa98      & $-$2.89 & 0.66\\
6362     & $-$1.15 &$-$0.58 & 14.67 & 0.09 & Wa99      & $-$2.85 & 0.66\\
\hline
\end{tabular}
\end{center}
* [Fe/H] value by Harris (1996). Ref. (RR Lyrae) - Wa94: Walker 1994;
SS95: Silbermann \& Smith 1995; Co99: Corwin et al. 1999; Ol99:
Olech et al. 1999; Wa96: Walker 1996; Ka01: Kaluzny et al. 2001;
LC99: Lee \& Carney 1999; Pi02: Piersimoni et al. 2002; CC01:
Corwin \& Carney 2001; Ka00: Kaluzny et al. 2000; Ca99: Caputo et
al. 1999; Wa98: Walker 1998; Wa99: Walker, private communication.
\end{table}

\begin{table}[h]
\begin{center}
\caption{$E(B-V)$ values and distance moduli (both in magnitudes)
for the selected clusters, as obtained from the various predicted
relations.} \vspace{0.3 cm}
\begin{tabular}{lcccccc}
\hline
NGC/IC &  $\mu_0$ & $E(B-V)$ & $E(B-V)$ & $\mu_V$ & $\mu_V$ \\
    &      $PWBV$&  $CA_{1.5}$ & $CA_{2.0}$& $PMA_{1.5}$ & $PMA_{2.0}$ \\
\hline
4590-M68 &15.08$\pm$0.07 & $-$0.01$\pm$0.02&+0.04$\pm$0.02 &15.05$\pm$0.10&15.24$\pm$0.10 \\
7078-M15 &15.13$\pm$0.08 & +0.02$\pm$0.02&+0.07$\pm$0.02  &15.23$\pm$0.08&15.41$\pm$0.08 \\
5466     &16.07$\pm$0.07 &$-$0.01$\pm$0.04&+0.04$\pm$0.03 &16.07$\pm$0.09&16.21$\pm$0.08 \\
6809-M55 &13.60$\pm$0.09 &+0.07$\pm$0.04&+0.12$\pm$0.04  &13.85$\pm$0.05&14.01$\pm$0.05 \\
4499     &16.45$\pm$0.08 &+0.16$\pm$0.04&+0.21$\pm$0.04 &17.01$\pm$0.08&17.17$\pm$0.08 \\
6934     &15.86$\pm$0.12 &+0.10$\pm$0.09&+0.15$\pm$0.09 &16.20$\pm$0.14&16.35$\pm$0.13 \\
7089-M2  &15.45$\pm$0.11 &$-$0.03$\pm$0.03&+0.02$\pm$0.03 &15.45$\pm$0.09&15.63$\pm$0.09 \\
3201     &13.28$\pm$0.08 &+0.24$\pm$0.04&+0.29$\pm$0.04 &14.10$\pm$0.13&14.28$\pm$0.12 \\
5272-M3  &15.06$\pm$0.07 &$-$0.03$\pm$0.02&+0.01$\pm$0.02 &15.00$\pm$0.08&15.18$\pm$0.07 \\
5904-M5  &14.27$\pm$0.16 &+0.00$\pm$0.03&+0.05$\pm$0.03 &14.36$\pm$0.09&14.54$\pm$0.09\\
1851     &15.42$\pm$0.13 &$-$0.03$\pm$0.04&+0.02$\pm$0.04 &15.38$\pm$0.17&15.56$\pm$0.17 \\
6362     &14.45$\pm$0.09 &+0.01$\pm$0.03&+0.06$\pm$0.03 &14.55$\pm$0.06&14.71$\pm$0.05 \\
\hline
\end{tabular}
\end{center}
\end{table}

\begin{table}[h]
\begin{center}
\caption{FOBE and FRE distance moduli (both in magnitudes) for the
selected clusters, as obtained with different assumptions on the
mixing-length parameter. The last column gives the adopted average
distance modulus.} \vspace{0.3 cm}
\begin{tabular}{lccccc}
\hline
 NGC&$\mu_V$(FOBE$_{1.5}$)& $\mu_V$(FRE$_{1.5}$)&
$\mu_V$(FOBE$_{2.0}$)& $\mu_V$(FRE$_{2.0}$) & $\langle \mu_V\rangle$\\
\hline
4590-M68&15.30$\pm$0.07&15.15$\pm$0.08&15.25$\pm$0.07&15.25$\pm$0.08&15.27$\pm$0.05\\
7078-M15&15.50$\pm$0.07&15.35$\pm$0.08&15.45$\pm$0.07&15.45$\pm$0.08&15.46$\pm$0.04\\
5466    &16.04$\pm$0.07&14.99$\pm$0.08&15.99$\pm$0.07&15.99$\pm$0.08&16.08$\pm$0.04\\
6809-M55&13.90$\pm$0.07&13.75$\pm$0.08&13.85$\pm$0.07&13.85$\pm$0.08&13.95$\pm$0.04\\
4499    &17.03$\pm$0.07&17.03$\pm$0.08&16.98$\pm$0.07&17.08$\pm$0.08 &17.09$\pm$0.04\\
6934    &16.19$\pm$0.07&16.19$\pm$0.08&16.14$\pm$0.07&16.24$\pm$0.08 &16.23$\pm$0.05\\
7089-M2 &15.43$\pm$0.07&15.43$\pm$0.08&15.38$\pm$0.07&15.48$\pm$0.08&15.50$\pm$0.04\\
3201    &14.25$\pm$0.07&14.10$\pm$0.08&14.20$\pm$0.07&14.20$\pm$0.08&14.24$\pm$0.05\\
5272-M3 &14.98$\pm$0.07&14.98$\pm$0.08&14.93$\pm$0.07&15.03$\pm$0.08&15.07$\pm$0.04\\
5904-M5 &14.34$\pm$0.07&14.34$\pm$0.08&14.29$\pm$0.07&14.39$\pm$0.08&14.41$\pm$0.04\\
1851    &15.39$\pm$0.07&15.39$\pm$0.08&15.34$\pm$0.07&15.44$\pm$0.08&15.42$\pm$0.05\\
6362    &14.52$\pm$0.07&14.52$\pm$0.08&14.47$\pm$0.07&14.57$\pm$0.08&14.63$\pm$0.04\\
\hline
\end{tabular}
\end{center}
\end{table}

\clearpage


\figcaption{Comparison between the computed period of models and
the value obtained using equation (1). The solid line is the
45$^o$ line.}


\figcaption{Period-Wesenheit ($PWBV$) relation for fundamental
(filled symbols) and fundamentalised first overtone (open symbols)
models. The solid line represents the linear regression through
the models [see equation (4)], while the dashed lines depict the
intrinsic dispersion.}


\figcaption{Comparison between the near-infrared magnitude of
models and the value obtained using equation (5). The solid line
is the 45$^o$ line, while the dashed lines depict the intrinsic
dispersion.}


\figcaption{Bolometric light curves with $l/H_p$=1.5 (solid line)
and $l/H_p$=2.0 (dashed line) for a number of F models with
$M=0.80{M_{\odot}}$ at varying luminosity and effective
temperature.}


\figcaption{The same but for FO models}


\figcaption{As in Fig. 2, but for the fundamental
Period-Magnitude-Amplitude ($PMA$) relations under the two values
of the mixing lenght parameter. The solid lines are the linear
regression through the data [see equations (6) and (7) in the
text), while the dashed lines depict the standard deviation.}


\figcaption{Upper panel: Visual amplitude $A_V$ as a function of
the intensity-weighted $\langle M_B\rangle-\langle M_V\rangle$
color for  F pulsators with $l/H_p$=1.5. The solid lines depict
the linear regression through the data [see equations (8a) and
(8b) in the text. Lower panel: As above, but with $l/H_p$=2.0. The
color index is corrected for the metal content and the solid line
is the linear regression trhough the data [see equation (9) in the
text.}


\figcaption{The average mass $\langle M(RR)\rangle$ (solar units)
of synthetic RR Lyrae pulsators versus the HB morphology, as
inferred from synthetic simulations with metal content $Z$=0.0001,
0.0003, 0.0006, 0.001, 0.003 and 0.006 (top to bottom). The
results for some selected globular clusters are reported (see
text).}


\figcaption{Color-magnitude diagram of globular cluster $RR_{ab}$
(filled symbols) and $RR_c$ (open symbols) variables, according to
intensity-averaged magnitudes scaled to an arbitrary value $V_r$
and $\langle B\rangle-\langle V\rangle$ colors corrected with the
$E(B-V)$ values in Table 4.}


\figcaption{Comparison between predicted and observed $PWBV$
relations (symbols as in Fig. 9). The period of $c$-type
 variables is fundamentalised and the solid
lines depict the predicted behavior at constant mass. The derived
intrinsic distance moduli are given in each panel.}


\figcaption{Comparison between our predicted intrinsic distance
moduli and those derived by Kovacs \& Walker (2001). The values
are scaled to the intrinsic distance modulus of IC 4499.}


\figcaption{Visual amplitude of cluster $ab$-type variables versus
the intrinsic $[\langle B\rangle-\langle V\rangle]_0$ color,
compared with the predicted relation at $l/H_p$=1.5 (solid line)
and $l/H_p$=2.0 (dashed line).}


\figcaption{As in Fig. 11, but for the $PMA$ relation of
fundamental pulsators at $l/H_p$=1.5. The solid lines are the
predicted slope at constant mass. The derived apparent distance
moduli are given in each panel.}


\figcaption{Observed distribution of RR Lyrae stars  in the
$M_V$-log$P$ plane in comparison with the predicted edges of the
instability strip at $l/H_p$=1.5 (solid lines). The distance
modulus labelled in each panel is obtained by fitting the observed
distribution of $c$-type variables (open symbols) to the predicted
FOBE. The dashed lines depict the observed limit of $ab$-type
variables under the hypothesis of a well populated instability
strip.}


\figcaption{As in Fig. 14, but adopting $l/H_p$=2.0.}


\figcaption{As in Fig. 11, but with the apparent distance moduli.}


\figcaption{Difference between intensity-weighted and
magnitude-weighted Wasenheit functions versus the visual
amplitude, as inferred by the pulsating models. The arrow depicts
the observed visual amplitude of RR Lyr.}

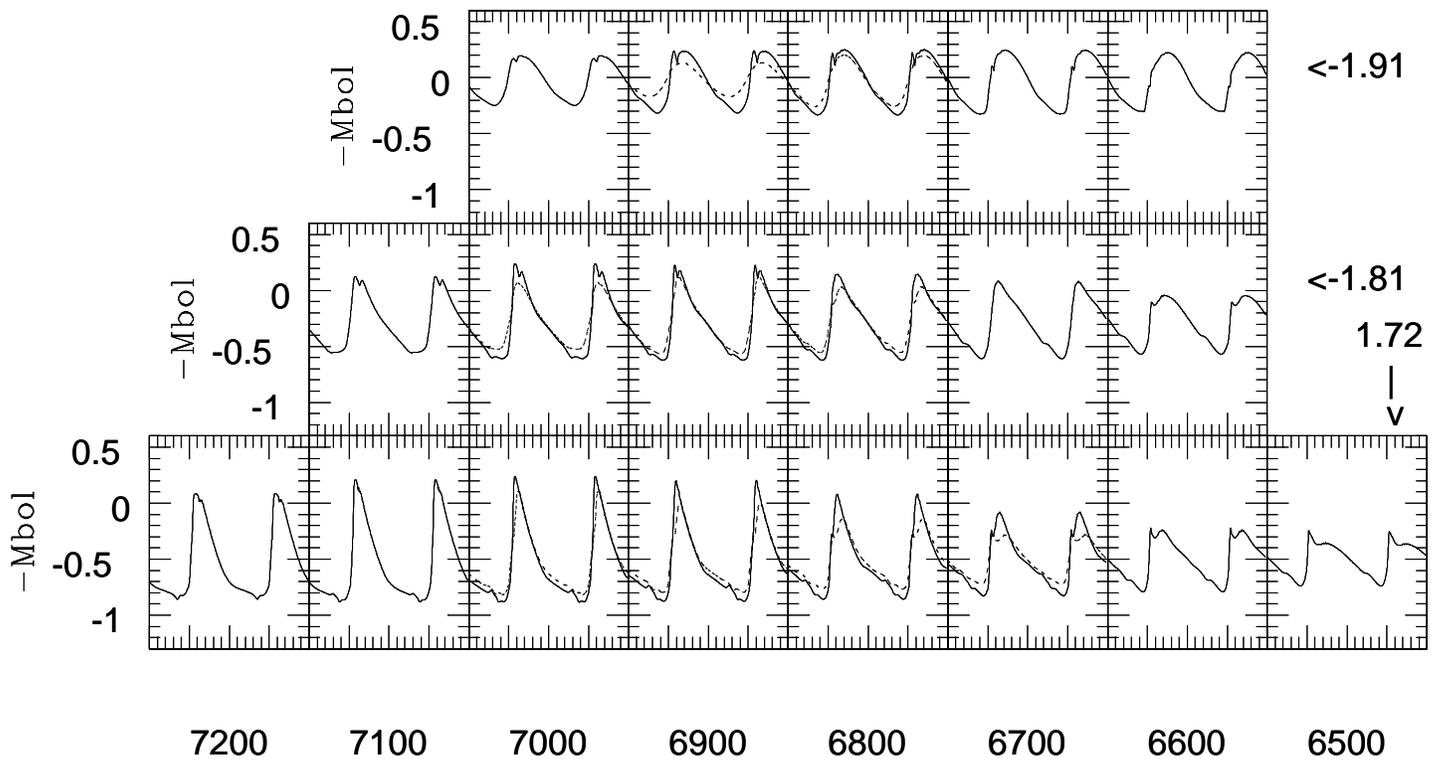
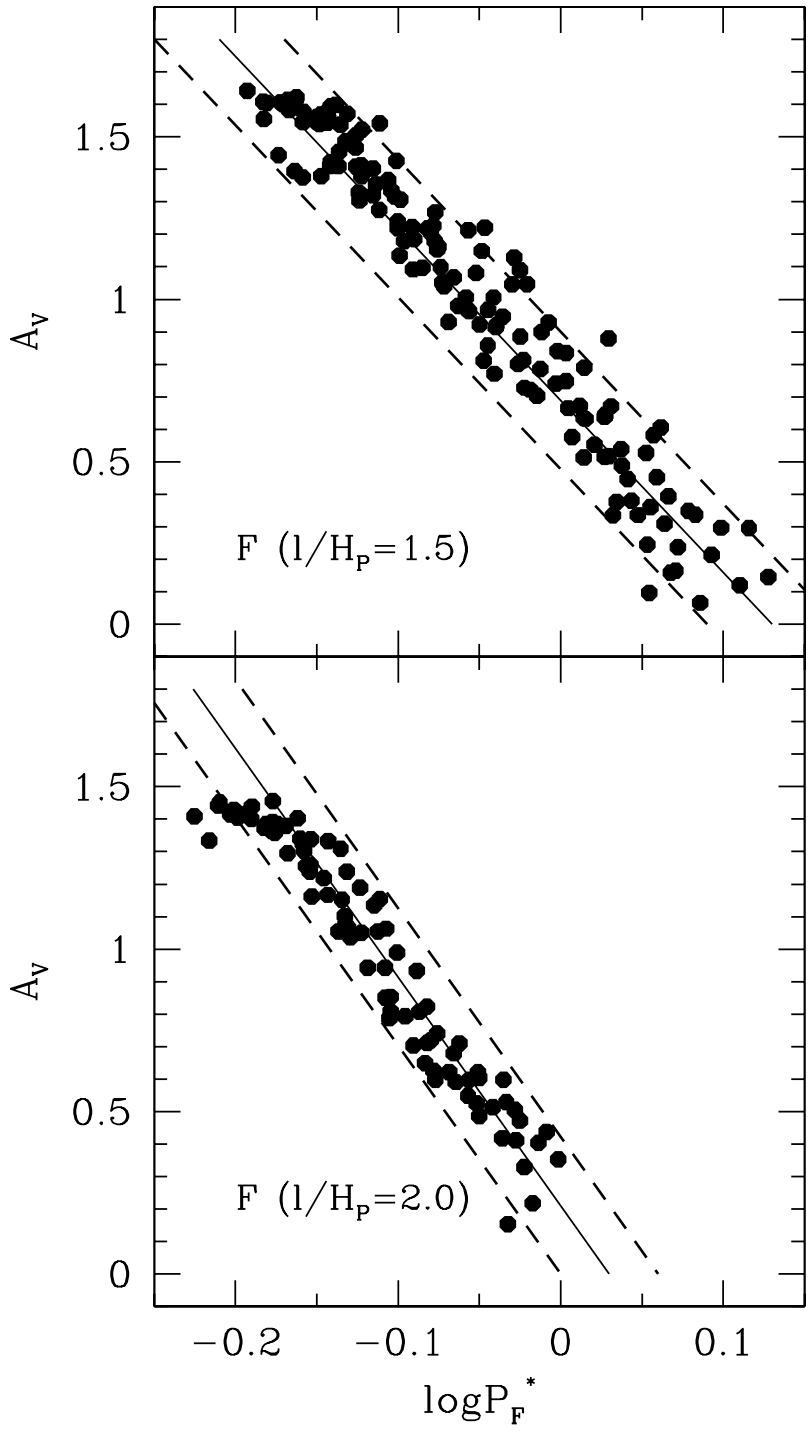
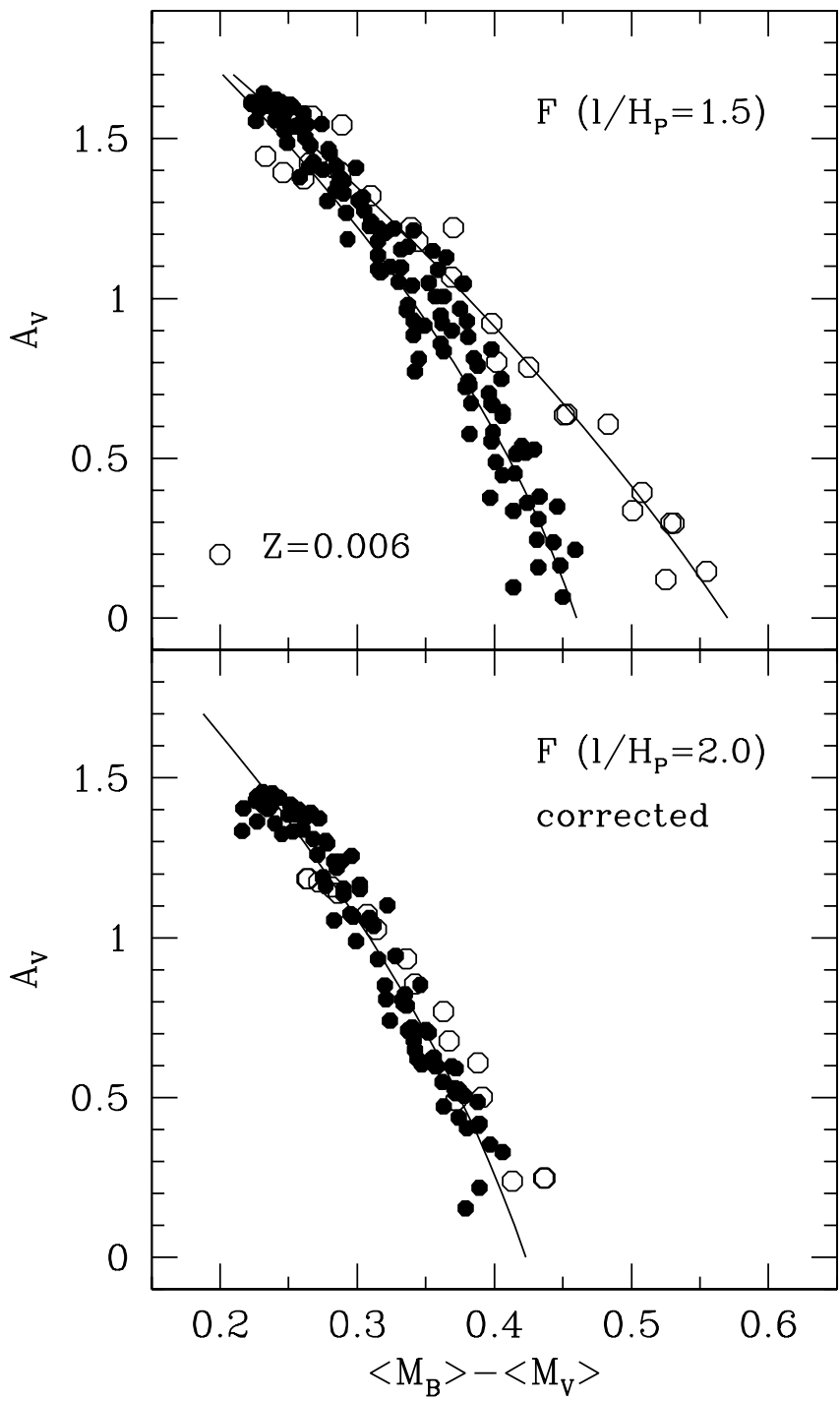
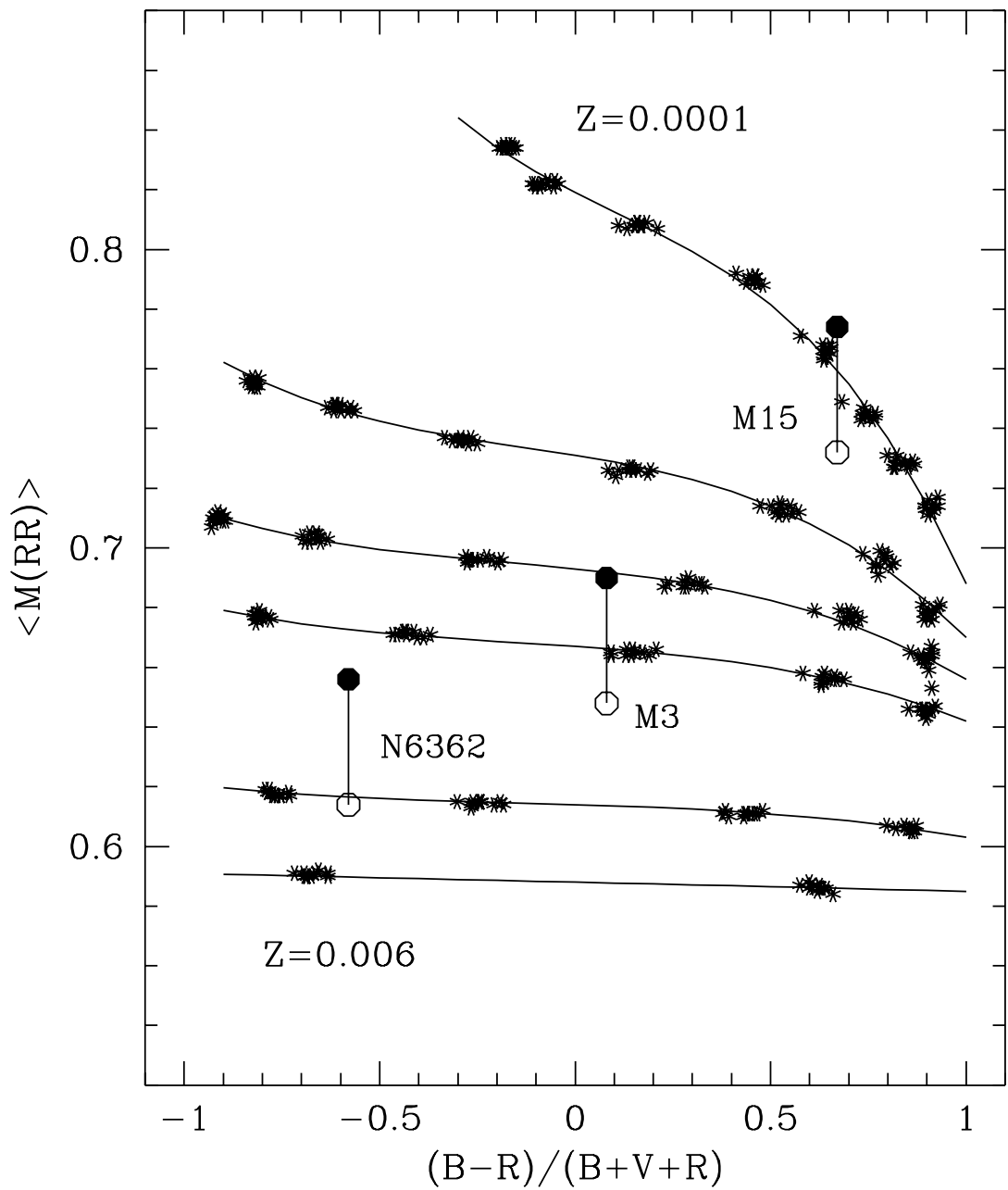
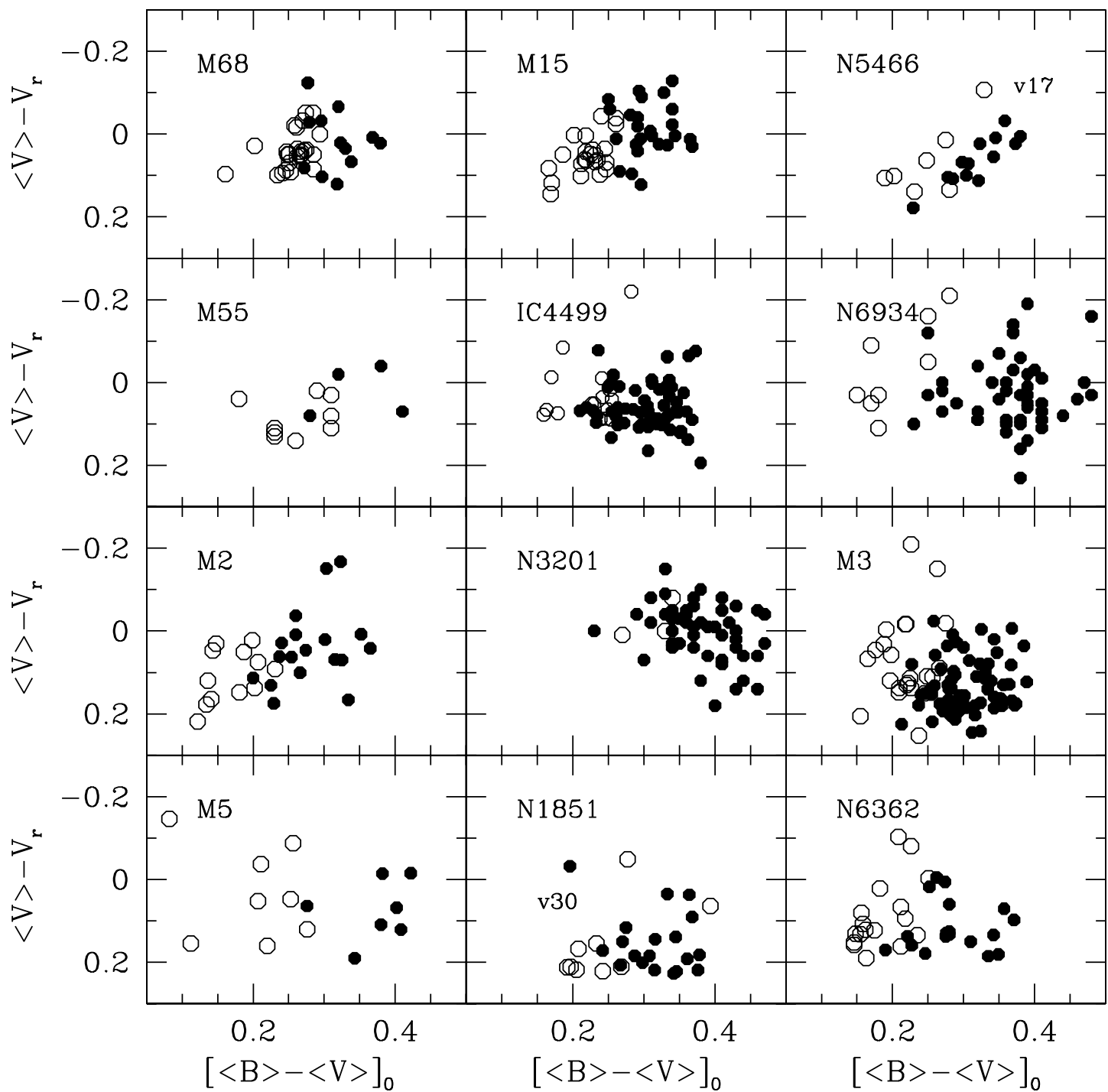
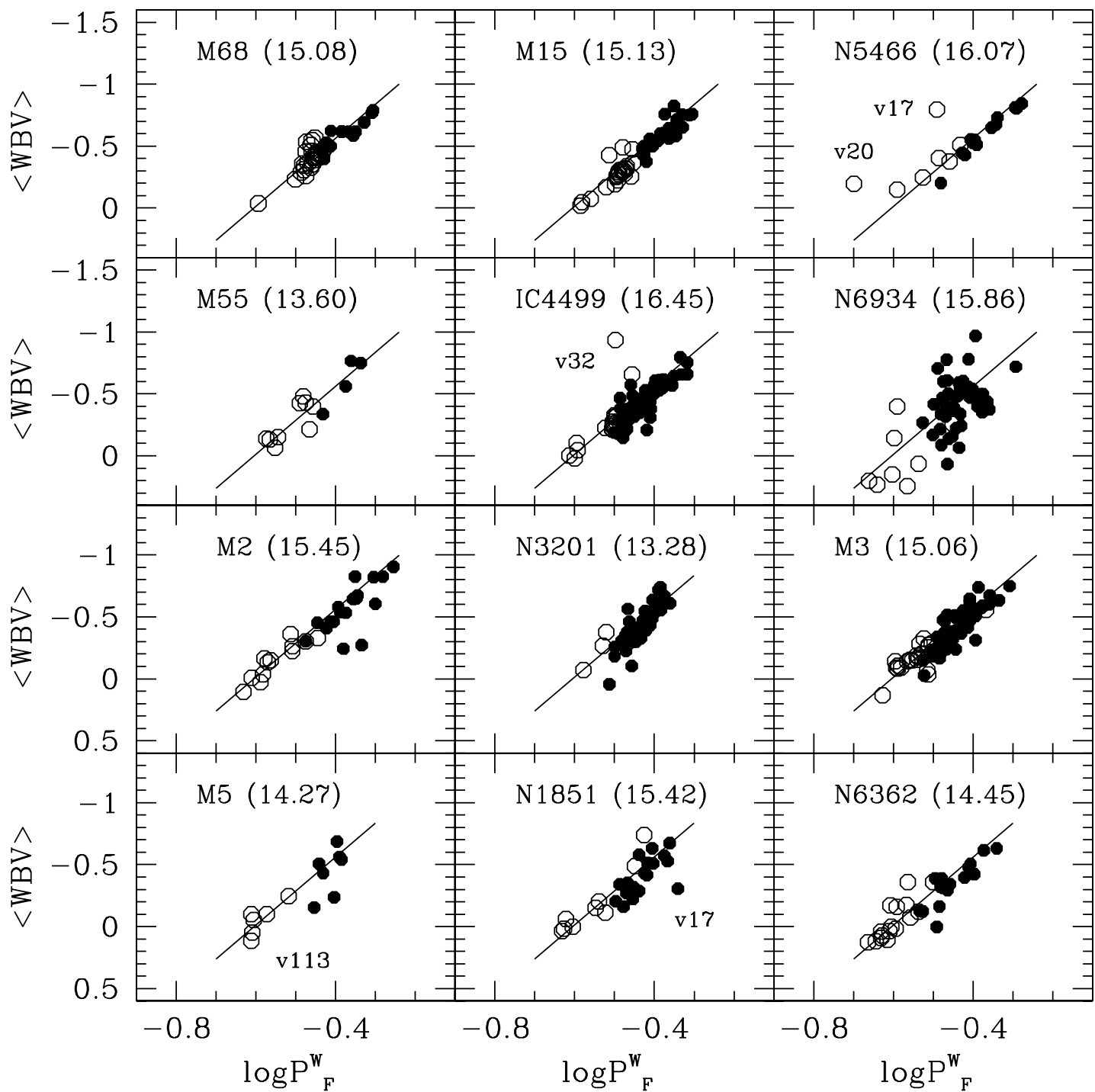
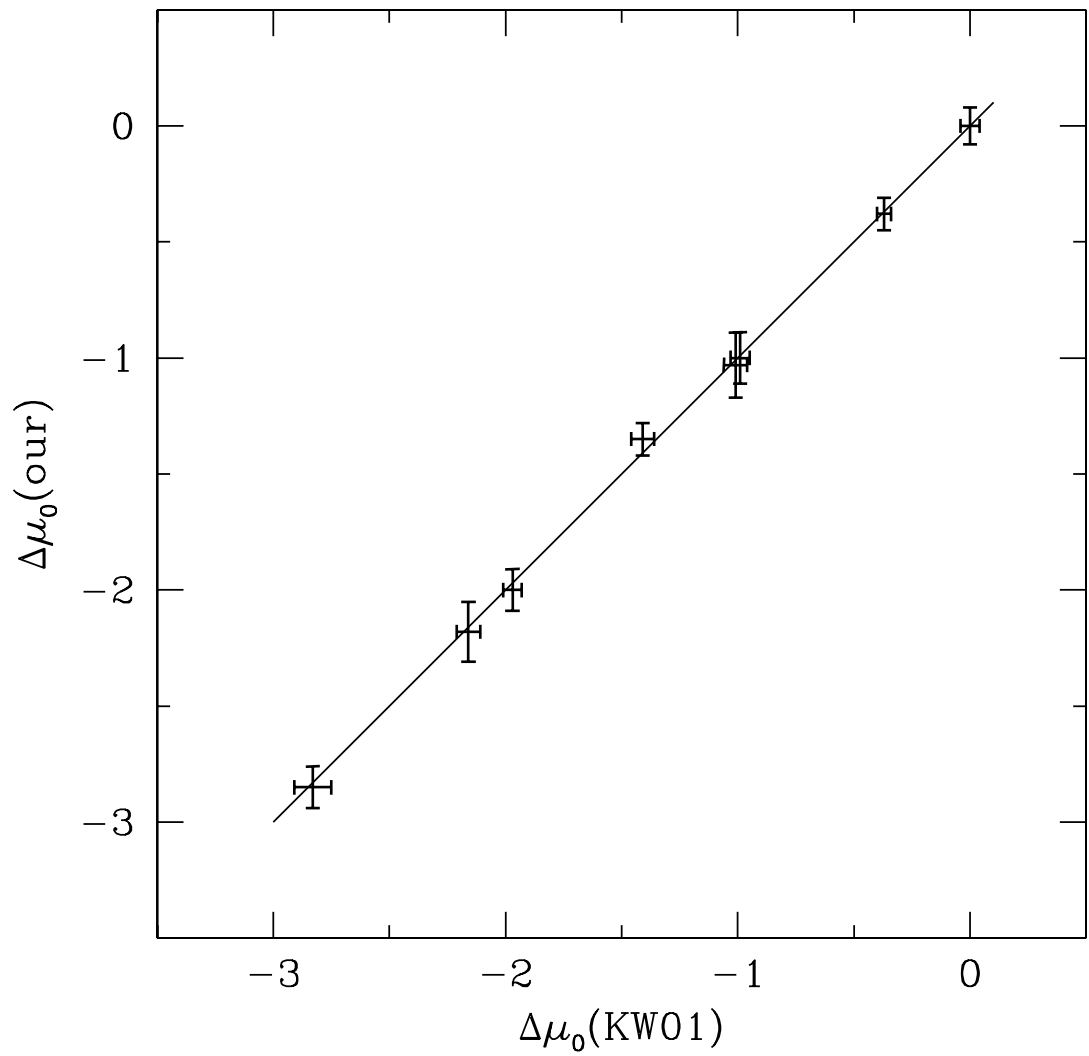
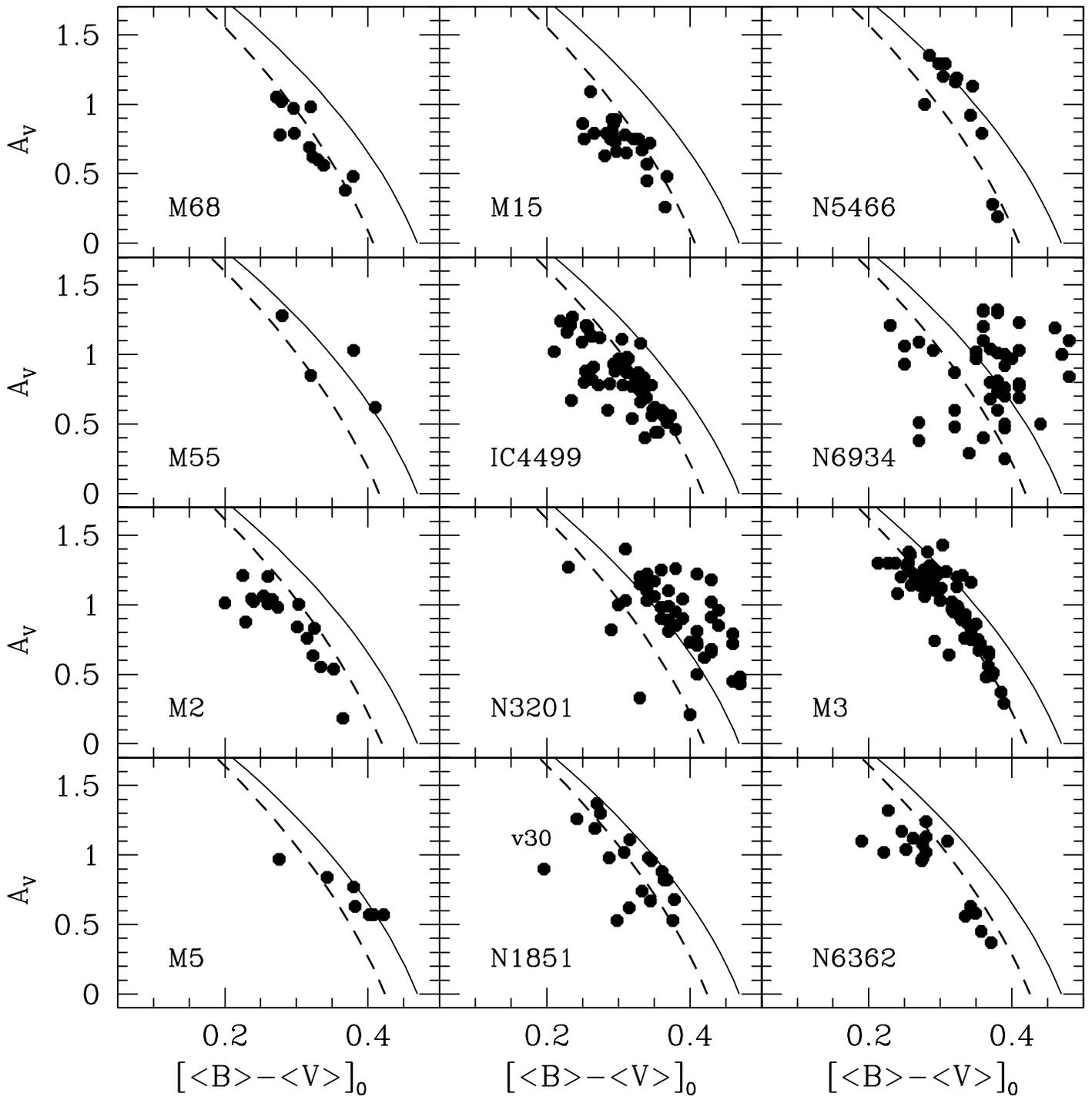
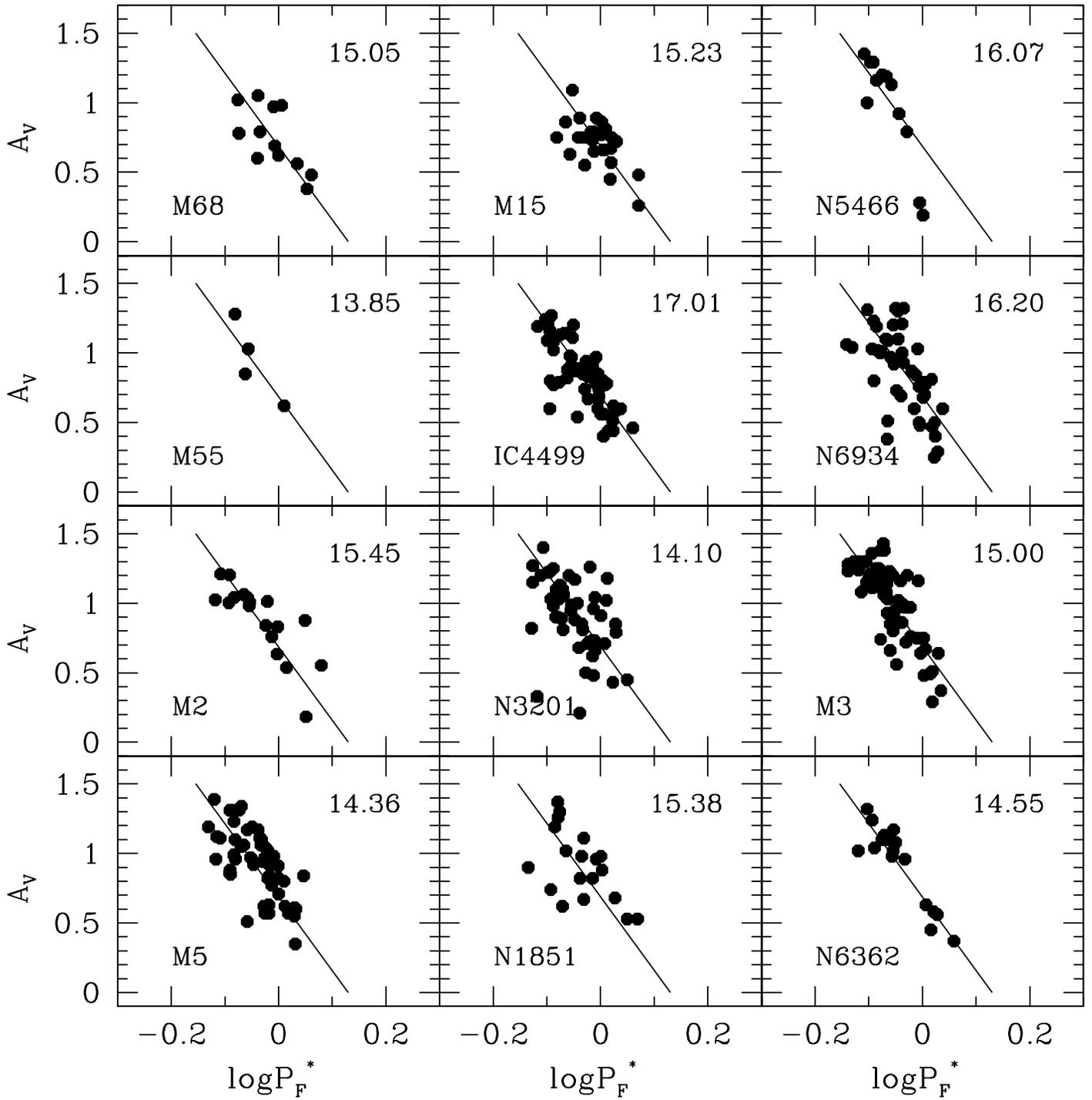
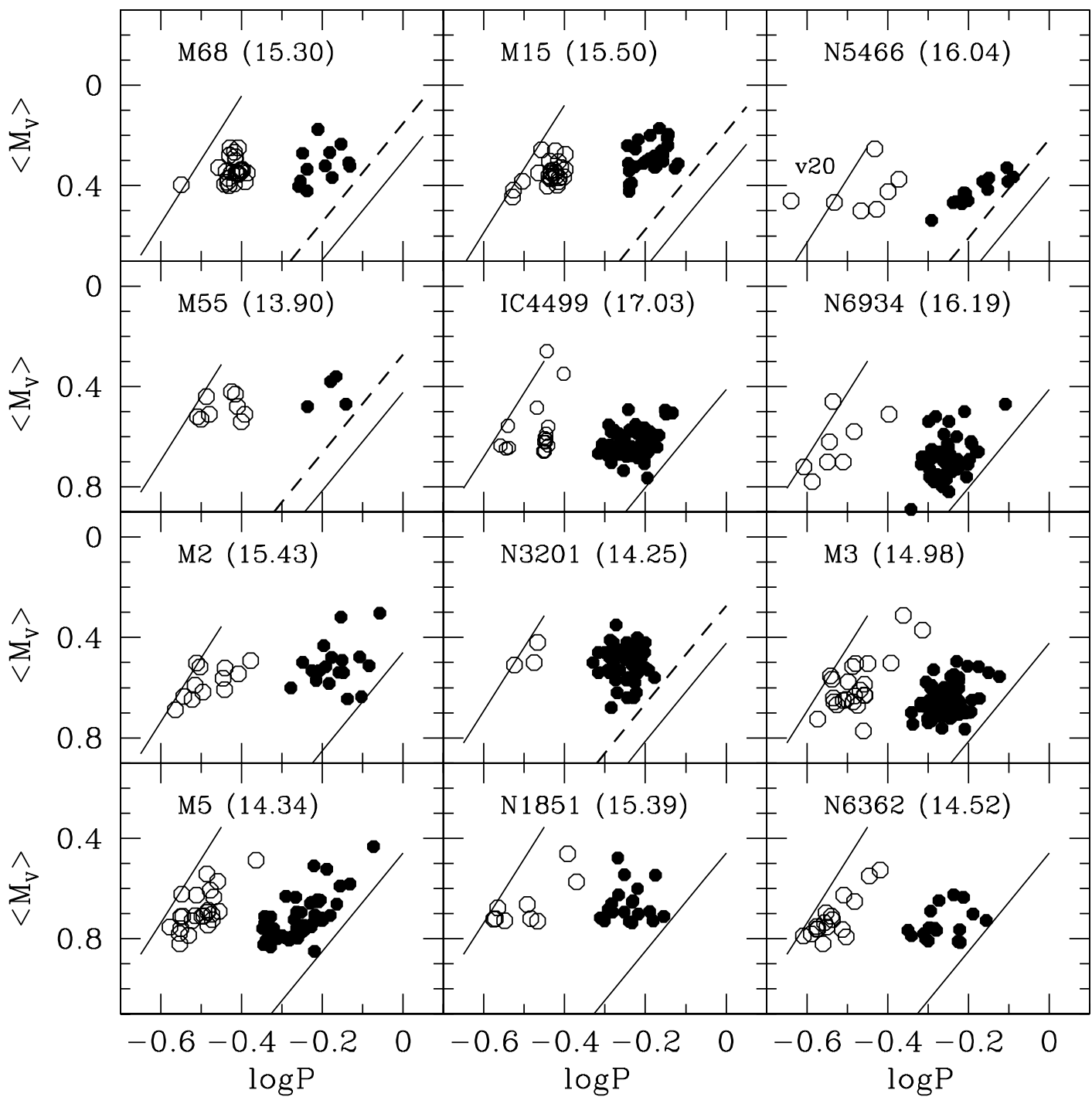
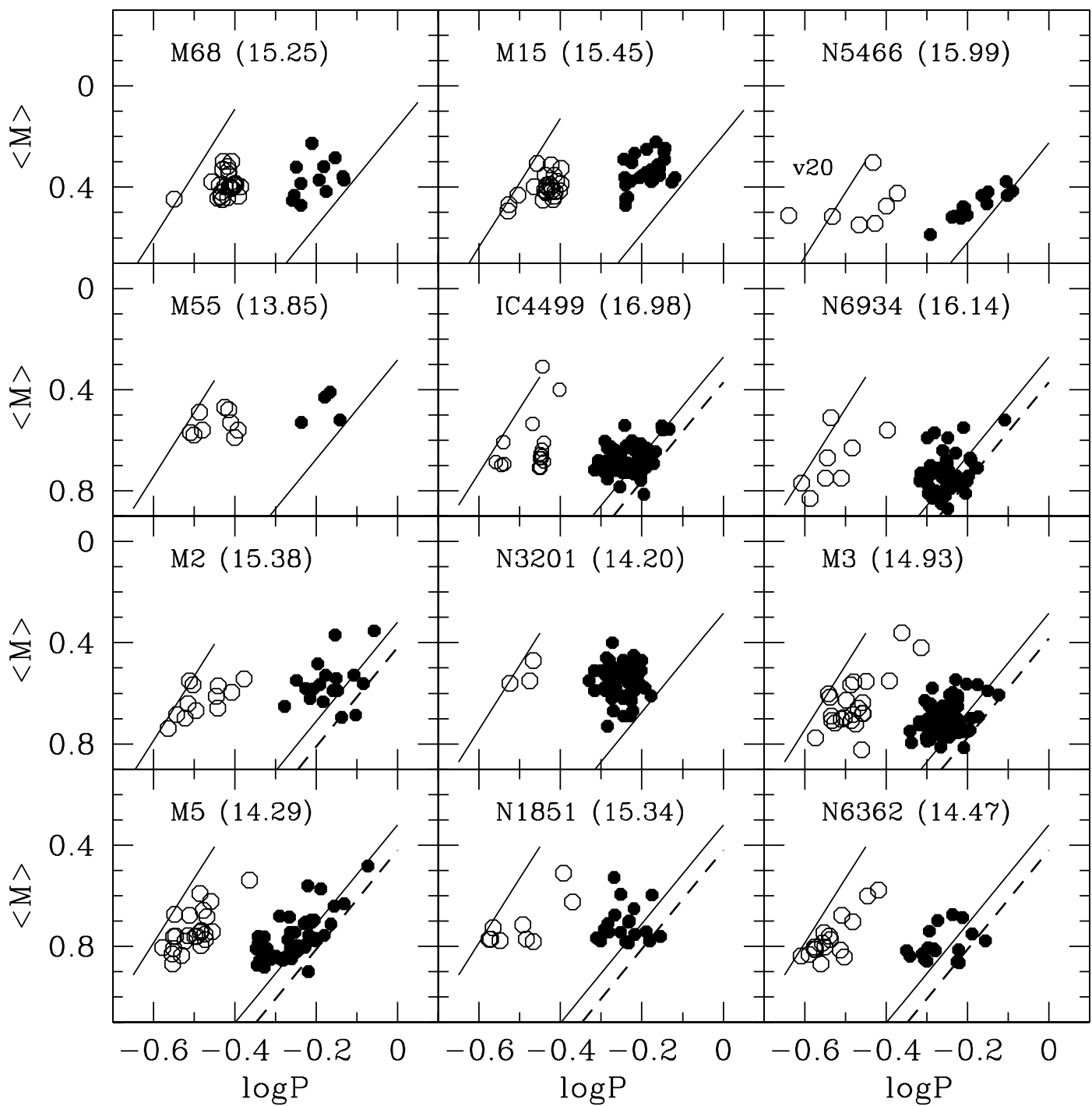
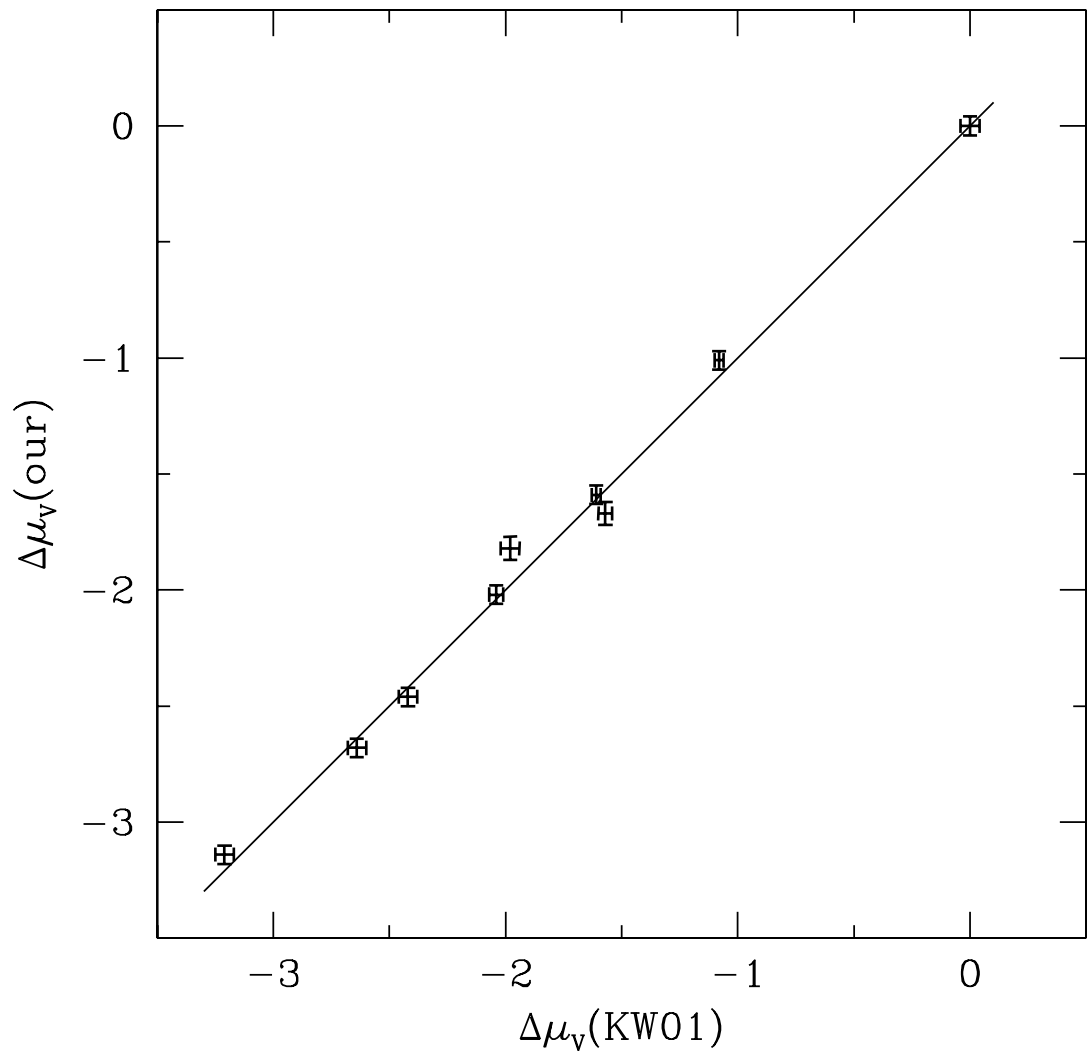
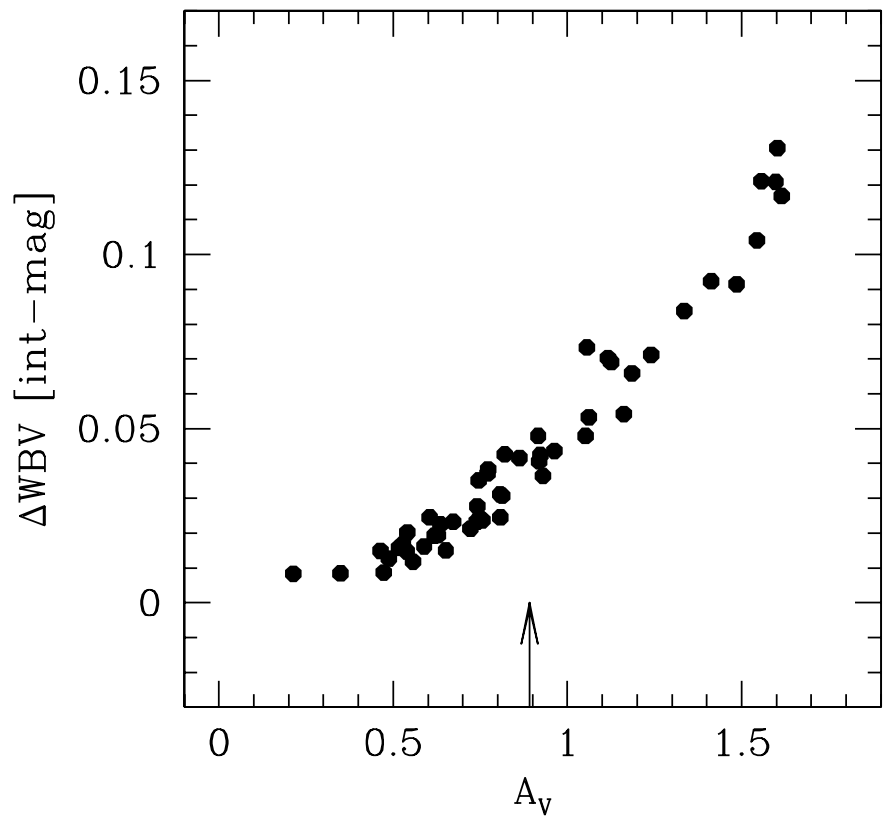
\end{document}